\documentclass[sn-mathphys-num]{sn-jnl}

\usepackage{graphicx}%
\usepackage{multirow}%
\usepackage{amsmath,amssymb,amsfonts}%
\usepackage{amsthm}%
\usepackage{mathrsfs}%
\usepackage[title]{appendix}%
\usepackage{xcolor}%
\usepackage{textcomp}%
\usepackage{manyfoot}%
\usepackage{booktabs}%
\usepackage{algorithm}%
\usepackage{algorithmicx}%
\usepackage{algpseudocode}%
\usepackage{listings}%

\usepackage{physics}
\newcommand{\ba}{\begin{array}}
\newcommand{\ea}{\end{array}}
\newcommand{\bmat}{\left(\begin{array}}
\newcommand{\emat}{\end{array}\right)}
\newcommand{\no}{\nonumber}

\newcommand{\be}{\begin{eqnarray}}
\newcommand{\ee}{\end{eqnarray}}

\theoremstyle{plain}
\newtheorem{theorem}{Theorem}
\newtheorem*{theorem*}{Theorem}

\newtheorem*{definition*}{Definition}
\newtheorem{lemma}[theorem]{Lemma}
\newtheorem*{lemma*}{Lemma}
\newtheorem{corollary}[theorem]{Corollary}
\newtheorem{remark}[theorem]{Remark}

\begin{document}

\title[Article Title]{Existence of long-range order in random-field Ising model on Dyson hierarchical lattice}


\author*[1]{\fnm{Manaka} \sur{Okuyama}}\email{manaka.okuyama.d2@tohoku.ac.jp}

\author[1,2,3]{\fnm{Masayuki} \sur{Ohzeki}}

\affil*[1]{\orgdiv{Graduate School of Information Sciences}, \orgname{Tohoku University}, \orgaddress{\city{Sendai}, \postcode{9808579}, \country{Japan}}}

\affil[2]{\orgdiv{Department of Physics}, \orgname{Institute of Science Tokyo}, \orgaddress{\city{Tokyo}, \postcode{1528551}, \country{Japan}}}

\affil[3]{\orgname{Sigma-i Co., Ltd.}, \orgaddress{\city{Tokyo}, \postcode{1080075}, \country{Japan}}}


\abstract{We study the random-field Ising model on a Dyson hierarchical lattice, where the interactions decay in a power-law-like form, $J(r)\sim r^{-\alpha}$, with respect to the distance. 
Without a random field, the Ising model on the Dyson hierarchical lattice has a long-range order at finite low temperatures when $1<\alpha<2$. 
In this study, for $1<\alpha<3/2$, we rigorously prove that there is a long-range order in the random-field Ising model on the Dyson hierarchical lattice at finite low temperatures, including zero temperature, when the strength of the random field is sufficiently small but nonzero. 
Our proof is based on Dyson's method for the case without a random field, and the concentration inequalities in probability theory enable us to evaluate the effect of a random field.
}

\keywords{Random-field Ising model, Dyson hierarchical lattice, long-range interaction}



\maketitle

\section{Introduction}\label{sec1}
Although there is no long-range order at a finite temperature in the one-dimensional short-range Ising model, a phase transition can occur in the one-dimensional long-range Ising model with a power-law decay interaction, $J(r)\propto r^{-\alpha}$, depending on the value of $\alpha$.
Dyson~\cite{Dyson} introduced the Ising model on a Dyson hierarchical lattice, to investigate the one-dimensional long-range Ising model. In this model, interactions decay in a power-law-like form $J(r)\sim r^{-\alpha}$ with respect to the distance, and the thermodynamic properties mimic those of the one-dimensional long-range Ising model.
Dyson first proved the existence of a long-range order at low temperatures in the Ising model on the Dyson hierarchical lattice for $1<\alpha<2$, and then proved the existence of a phase transition in the one-dimensional long-range Ising model for $1<\alpha<2$ using the Griffiths--Kelly--Sherman inequality~\cite{Griffiths,KS}.
For $\alpha=2$, the existence of a phase transition in the one-dimensional long-range Ising model was proven using contours by Fr\"{o}hlich and Spencer~\cite{FS}.
Mori~\cite{Mori} proved that, for $0\le\alpha<1$ with appropriate normalization, the free energy of the one-dimensional long-range Ising model coincides exactly with that of the corresponding mean-field model.
Additionally, there is no phase transition for $2<\alpha$~\cite{Dobruschin,Dobruschin2,Ruelle}.

The one-dimensional long-range Ising model can be simply extended to a random system by considering a random field.
In the one-dimensional long-range random-field Ising model, Aizenman and Wehr~\cite{AW} proved that no phase transition occurs when $3/2<\alpha$.
Tsuda and Nishimori~\cite{TN} proved that, for $0\le\alpha<1$ with appropriate normalization, the free energy of the one-dimensional long-range random-field Ising model is equivalent to that of the corresponding mean-field model.
On the other hand, Imry-Ma argument~\cite{IM} suggested the existence of a phase transition for $1<\alpha<3/2$, and Cassandro, Orlandi, and Picco~\cite{COP} provided a rigorous proof using contours for $3- \log 3/\log 2<\alpha<3/2$.
However, there is no rigorous result for $1<\alpha\le3- \log 3/\log 2$, although considerable evidence supports the existence of a phase transition in this interval~\cite{Bray,WRK,MG,LP,BTT}

Similar to the case without a random field, the random-field Ising model on the Dyson hierarchical lattice was considered to mimic the properties of a one-dimensional long-range random-field Ising model.
Various studies~\cite{RB,BTT,MG,PR,DPR} have supported the existence of a long-range order in the random-field Ising model on the Dyson hierarchical lattice for $1<\alpha<3/2$.
However, a rigorous proof is yet to be provided.

In this study, for $1<\alpha<3/2$, we rigorously prove the existence of a long-range order in the random-field Ising model on the Dyson hierarchical lattice at finite low temperatures when the strength of the random field is sufficiently small.
Our proof is based on the case without a random field, where Dyson~\cite{Dyson} rigorously proved the existence of a long-range order by ingeniously utilizing a hierarchical structure and deriving a recurrence formula for the square of the total magnetization in a finite system.
At first glance, Dyson's method appears inapplicable in the presence of random fields.
However, we rigorously evaluated the contribution of random fields in a finite system using concentration inequalities~\cite{BLM} and successfully extended Dyson's method to the case of random fields.

Unfortunately, our result does not guarantee the existence of a long-range order in the one-dimensional long-range random-field Ising model for $1<\alpha\le3- \log 3/\log 2$ because the Griffiths--Kelly--Sherman inequality does not hold, unlike the case without a random field.
Nevertheless, as the random-field Ising model on the Dyson hierarchical lattice is expected to mimic the properties of the one-dimensional long-range random-field Ising model, our results strongly support the existence of a phase transition in the one-dimensional long-range random-field Ising model for $1<\alpha\le3- \log 3/\log 2$.

The remainder of this paper is organized as follows.
In Sec. II, we define the random-field Ising model on the Dyson hierarchical lattice and present the main results.
In Sec. III, we provide the proof of the theorem.
In Sec. IV, we discuss the results.

\section{Model and result}\label{sec2}
For each positive integer $N$, there are $2^N$ Ising spins $\sigma_i=\pm1$ labeled by the index $i=1,2,\cdots,2^N$.
The following recurrence relation defines the Ising model for a Dyson hierarchical lattice
\be
H_{N}(\vec{\sigma})
&=&H_{N-1}(\vec{\sigma}_1) + H_{N-1}(\vec{\sigma}_2) -\frac{b_{N}}{2^{2 N}} \sum_{i,j=1}^{2^{N}} \sigma_i \sigma_ j  
\no\\
&=&H_{N-1}(\vec{\sigma}_1) + H_{N-1}(\vec{\sigma}_2) -\frac{1}{2^{\alpha N}}\sum_{i,j=1}^{2^{N}} \sigma_i \sigma_ j  , \label{Dyson-H}
\\
H_0(\vec{\sigma})&=&0 ,
\ee
where $b_{N}=2^{(2-\alpha)N}$, $1<\alpha$, $\sigma_i=\pm1$, $\vec{\sigma}_1\equiv \{ \sigma_i \}_{1\le i \le 2^{N-1}}$, and $\vec{\sigma}_2\equiv \{ \sigma_i \}_{2^{N-1}+1\le i \le 2^{N}}$.
The thermodynamic limit of the free energy exists when $\alpha>1$ (for example, see Ref. \cite{GM,CBG} for the proof).

The random-field Ising model on the Dyson hierarchical lattice is defined as
\be
H_{N}^{\mathrm{RF}}(\vec{\sigma})&=&H_{N}(\vec{\sigma}) - h\sum_{i=1}^{2^N} h_i \sigma_i,
\ee
where  $h>0$ is the strength of random fields $h_i$, and we consider the following two probability distribution cases of $h_i$: (i) a Gaussian distribution $\mathcal{N}(0,1)$ and (ii) a symmetric Bernoulli distribution $\Pr(h_i=1)=\Pr(h_i=-1)=1/2$.
See Remark \ref{remark3} for the case of more general probability distributions.

The thermal average with respect to $H_{N}^{\mathrm{RF}}(\vec{\sigma})$ is defined as
\be
\langle \cdots\rangle_N&=&\frac{\Tr( \cdots e^{-\beta H_{N}^{\mathrm{RF}}(\vec{\sigma})})}{\Tr( e^{-\beta H_{N}^{\mathrm{RF}}(\vec{\sigma})})},
\ee
where $\beta$ is the inverse temperature and $\Tr$ denotes the summation with respect to all the spin variables.
The total spin $S_{N,1}$ is given by
\be
S_{N,1}&=& S_{N-1,1}+S_{N-1,2},
\\
S_{N-1,1}&=&\sum_{i=1}^{2^{N-1}}\sigma_i \label{Spr},
\\
S_{N-1,2}&=&\sum_{i=2^{N-1}+1}^{2^{N}}\sigma_i .
\ee
We define $f_N$ as the quenched average of the square of the total magnetization
\be
f_N&=&\frac{1}{2^{2N}} \mathbb{E}[ \langle S_{N,1}^2  \rangle_N], \label{def-fN}
\ee
where $\mathbb{E}\left[\cdots\right]$ denotes the expectation with respect to all the random variables.
We note that the existence of the thermodynamic limit, $\lim_{N\to\infty}f_N$, is not guaranteed.
Instead, we define the long-range order parameter in the thermodynamic limit~\cite{Dyson} as
\be
m^2  &\equiv&\liminf_{N\to\infty}f_N.
\ee
From this definition, $m^2$ always exists, and $m^2\neq0$ indicates that the spins are correlated at arbitrarily large distances.
Thus, $m^2$ plays a role in the long-range order parameters.

For both Gaussian and Bernoulli distributions of the random fields $h_i$, our main result is the same lower bound on $m^2$:
\begin{theorem}[]\label{theorem1}
For $1<\alpha <3/2$ and $1\le(c 2^{(2-\alpha)})^{1/2}$, the long-range order parameter is bounded by
\be
m^2
&\ge&  1- \frac{1}{c}\frac{2^\alpha}{2^2-2^\alpha} - h  \left( \frac{(2-\alpha) 2^{1/2+\alpha}}{(2^{3/2}-2^\alpha)^2}\log  2 + \frac{2^\alpha}{2^{3/2}-2^\alpha}\log( 2c ^{1/2}(1+\sqrt{2\pi e}   )) \right)
\no\\
&&- \frac{1}{\beta} \left(\frac{2^\alpha}{2^2-2^\alpha} \log (2c^{1/2}) + \frac{(2-\alpha) 2^{1+\alpha}}{(2^{2}-2^\alpha)^2}\log  2\right). \label{m-bound}
\ee
\end{theorem}
As the right-hand side of Eq. (\ref{m-bound}) decreases monotonically with respect to $\beta$,  there will always be a long-range order at zero temperature if $m^2>0$ at a finite low temperature.
By choosing $c$ and $\beta$ sufficiently large and $h$ sufficiently small, it is possible to show $m^2>0$ with $1<\alpha <3/2$ when $h\neq0$ (see Fig. 1 for example).
Thus, we arrive at the following conclusions:

\begin{corollary}[]
For $1<\alpha <3/2$, the random-field Ising model on the Dyson hierarchical lattice has a long-range order at finite low temperatures, including zero temperature, when the strength of the random fields is sufficiently small.
\end{corollary}
\begin{figure}[H]
\centering
\includegraphics[height=0.45\columnwidth]{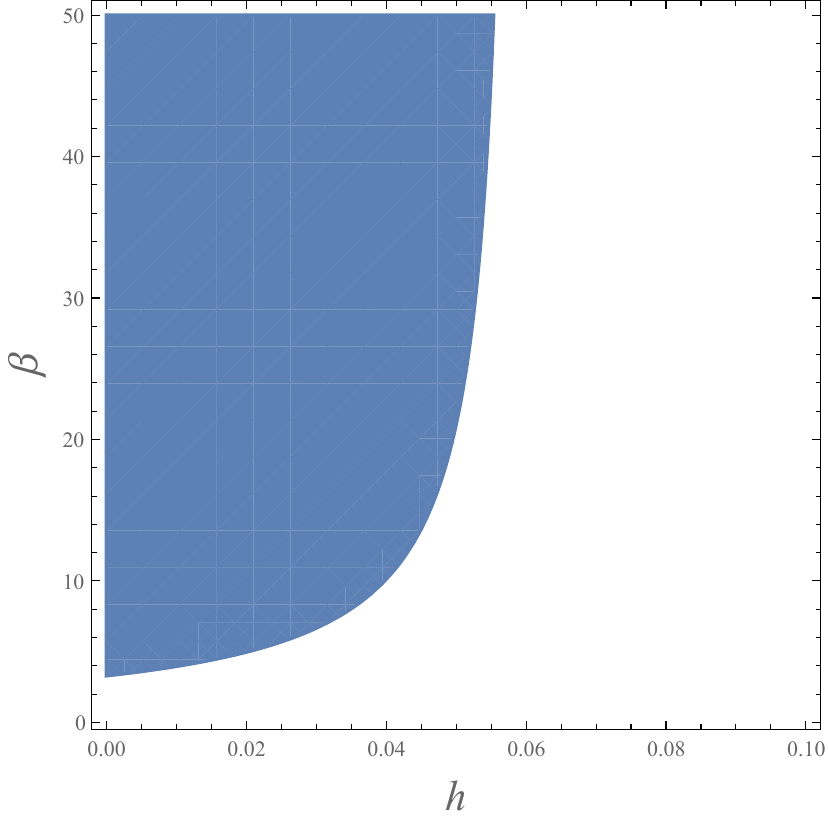}
\includegraphics[height=0.45\columnwidth]{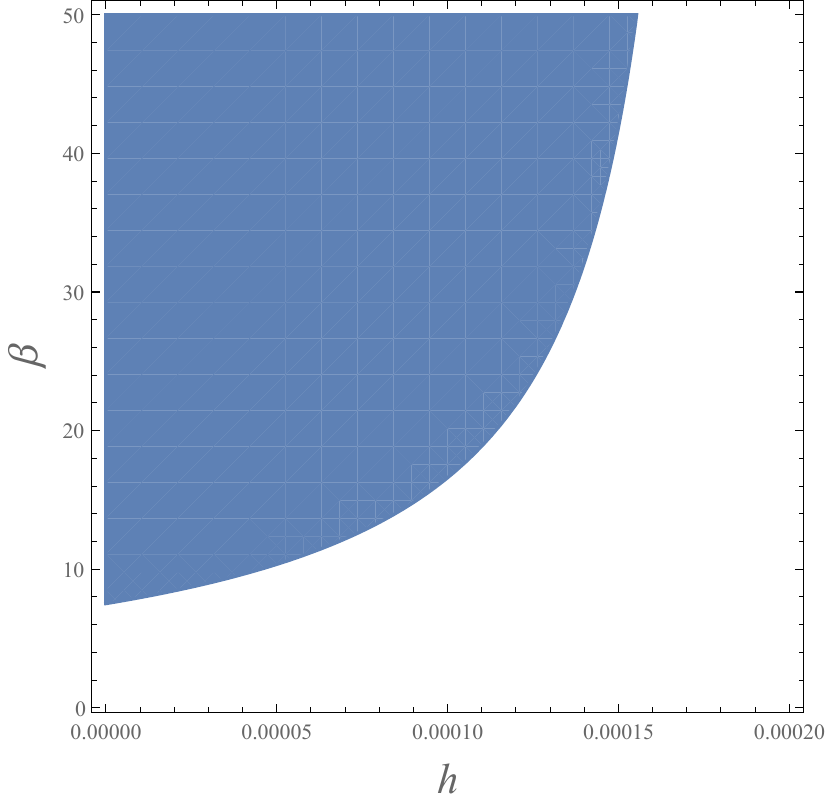}\\
\caption{(Color online) 
Regions where $m^2$ is greater than 0 examined numerically based on Eq. (\ref{m-bound}).
The horizontal and vertical axes denote the strength of the random field and the inverse temperature, respectively.
Left panel: For $\alpha=1.1$ and $c=10$, the blue region guarantees $m^2>0$.
Right pane: For $\alpha=1.49$ and $c=10$, the blue region guarantees $m^2>0$.}
\label{fig1}
\end{figure}
\begin{remark}\label{remark3}
Theorem \ref{theorem1} can be extended to random fields following continuous or discrete i.i.d. symmetric random variables beyond the Gaussian and Bernoulli cases by using the modified logarithmic Sobolev inequality under suitable conditions on the probability distribution (See chapter 6 in Ref. \cite{BLM} for details).
\end{remark}

\section{Proof of Theorem \ref{theorem1}}\label{sec3}
The proofing strategy involves deriving a recursive inequality between $f_N$ and $f_{N-1}$~\cite{Dyson}.

Equation (\ref{Spr}) implies
\be
-2^{(N-1)}\le S_{N-1,1}  \le 2^{(N-1)} \label{interval}.
\ee
For $c>0$ and $d\ge0$, let $r_N$ be a positive integer such that
\be
(c \beta^d b_N)^{1/2}< r_N \le 1+(c \beta^d b_N)^{1/2}, \label{lange-r_N}
\ee
and then, the interval (\ref{interval}) is divided into $r_N$ equal parts, $I_k$ $(k=1,2,\cdots, r_N)$.
For $S_{N-1,1}, S_{N-1,2}\in I_k $, we have
\be
(S_{N-1,1} -S_{N-1,2})^2 \le \frac{2^{2N}}{r_N^2}<\frac{2^{2N}}{c \beta^d b_N}. \label{res-S1-S2}
\ee
We define the restricted partition function as
\be
Z_{k,1}(\vec{h}_1)&\equiv&\Tr_{ \{ S_{N-1,1}\in I_k\}}  \exp(-\beta H_{N-1}^{\mathrm{RF}}(\vec{\sigma}_1) +   2 \beta \frac{ b_N }{2^{2 N}}  S_{N-1,1}^2),
\\
Z_{k,2}(\vec{h}_2)&\equiv&\Tr_{ \{ S_{N-1,2}\in I_k\}}  \exp(-\beta H_{N-1}^{\mathrm{RF}}(\vec{\sigma}_2)    + 2 \beta \frac{ b_N }{2^{2 N}} S_{N-1,2}^2  ),
\ee
where  $\vec{h}_1\equiv \{ h_i \}_{1\le i \le 2^{N-1}}$, $\vec{h}_2\equiv \{ h_i \}_{2^{N-1}+1\le i \le 2^{N}}$, and $\Tr_{ \{ S_{N-1,1}\in I_k\}}$ and $\Tr_{ \{ S_{N-1,2}\in I_k\}}$ denote the summation for the spin variables $\vec{\sigma}_1\equiv \{ \sigma_i \}_{1\le i \le 2^{N-1}}$ and $\vec{\sigma}_2\equiv \{ \sigma_i \}_{2^{N-1}+1\le i \le 2^{N}}$, respectively, where the values of $S_{N-1,1}$ and $S_{N-1,2}$ are restricted to the interval $I_k$.
Note that, for $h=0$, $Z_{k,1}(\vec{h}_1) =Z_{k,2}(\vec{h}_2)$, but for $h\neq0$, $Z_{k,1}(\vec{h}_1) \neq Z_{k,2}(\vec{h}_2)$.

Then, by using the Gibbs-Bogoliubov inequality~\cite{Kuzemsky}, we obtain the following recurrence relation:
\begin{lemma}[] \label{lemma3}
The quenched average of the square of the total magnetization satisfies the following recurrence inequality
\be
f_N
&\ge&f_{N-1} -\frac{1}{\beta b_N}\left( \frac{1}{c \beta^{d-1}} +\log (1+(c \beta^d b_N)^{1/2}) +\mathbb{E}[\max _k \log ( \frac{Z_{k,1}(\vec{h}_1)  } { Z_{k,2}(\vec{h}_2) })] \right). \label{bottle} 
\no\\\ee\
\end{lemma}
\begin{remark}
When there is no random field ($h=0$), by setting $c=1$ and $d=1$, Eq. (\ref{bottle}) is deduced to Eq. (4.25) in Ref. \cite{Dyson}
\be
f_N
&\ge&f_{N-1} -\frac{1}{\beta b_N}\left( 1 +\log (1+( \beta b_N)^{1/2})  \right).
\ee
Then, we can prove the existence of a long-range order in the Ising model on the Dyson hierarchical lattice when $0<\alpha<1$, as shown in Ref. \cite{Dyson}.
\end{remark}
\begin{proof}[\textbf{Proof of Lemma \ref{lemma3}}]
First, we define the quenched pressure function of the random-field Ising model on the Dyson hierarchical lattice as follows:
\be
P_N(v_N,v_{N-1}.\cdots,v_1)&=& \mathbb{E}[\log\Tr  e^{-\beta H_{N}^{\mathrm{RF}}(\vec{\sigma}) } ]
=\mathbb{E}[\log\Tr  e^{-\beta H_{N}^{}(\vec{\sigma} )  +\beta h\sum_{i=1}^{2^N} h_i \sigma_i} ]
\no\\
&=&\mathbb{E}[\log\Tr  e^{-\beta H_{N-1}(\vec{\sigma}_1) -\beta H_{N-1}(\vec{\sigma}_2) +v_N\sum_{i,j=1}^{2^{N}} \sigma_i \sigma_ j   +\beta h\sum_{i=1}^{2^N} h_i \sigma_i} ],
\no\\
\ee
where $v_N=\beta b_N /2^{2 N}$ and $\Tr$ denotes the summation of all the spin variables.
The Gibbs-Bogoliubov inequality~\cite{Kuzemsky} indicates that
\be
P_N(v_N,v_{N-1},\cdots,v_1)&\le& P_N(0,v_{N-1},\cdots,v_1)  +   v_N \mathbb{E}[\langle S_{N,1}^2 \rangle_N]
\no\\
&=& 2P_{N-1}(v_{N-1},\cdots,v_1)  +   v_N 2^{2N}f_N,
\\
2P_{N-1}(2v_N+v_{N-1},v_{N-2},\cdots,v_1) 
&\ge&2P_{N-1}(v_{N-1},v_{N-2},\cdots,v_1) + 4 v_N \mathbb{E}[\langle S_{N-1,1}^2 \rangle_{N-1}] 
\no\\
&=&2P_{N-1}(v_{N-1},v_{N-2},\cdots,v_1) +  v_N 2^{2N} f_{N-1}. 
\ee
Thus, we obtain
\be
f_N-f_{N-1}&\ge&\frac{1}{\beta b_N}\left(P_N(v_N,v_{N-1},\cdots,v_1)-2P_{N-1}(2v_N+v_{N-1},v_{N-2},\cdots,v_1) \right).
\no\\\label{f-P}
\ee

On the other hand,  by using the restricted summation $\Tr_{ \{ S_{N-1,1}, S_{N-1,2}\in I_k\}}$ where the values of $S_{N-1,1}$ and $S_{N-1,2}$ are restricted to the interval $I_k$, we have
\be
&&P_N(v_N,v_{N-1},\cdots,v_1)
\no\\
&=&\mathbb{E}[\log\Tr  e^{-\beta H_{N-1}^{}(\vec{\sigma}_1 )+ 2 v_N S_{N-1,1}^2-\beta H_{N-1}^{}(\vec{\sigma}_2 ) +  2 v_NS_{N-1,2}^2  +\beta h\sum_{i=1}^{2^N} h_i \sigma_i -v_N(S_{N-1,1}-S_{N-1,2})^2} ]
\no\\
&\ge&\mathbb{E}[\log\sum_{k=1}^{r_N} \Tr_{ \{ S_{N-1,1},S_{N-1,2}\in I_k\}}   e^{-\beta H_{N-1}^{}(\vec{\sigma}_1 )+ 2 v_N S_{N-1,1}^2-\beta H_{N-1}^{}(\vec{\sigma}_2 ) +  2 v_N S_{N-1,2}^2 }
\no\\
&&e^{ \beta h\sum_{i=1}^{2^N} h_i \sigma_i -v_N(S_{N-1,1}-S_{N-1,2})^2} ]
\no\\
&\ge&\mathbb{E}[\log\sum_{k=1}^{r_N} \Tr_{ \{ S_{N-1,1},S_{N-1,2}\in I_k\}}   e^{-\beta H_{N-1}^{}(\vec{\sigma}_1 )+ 2 v_NS_{N-1,1}^2-\beta H_{N-1}^{}(\vec{\sigma}_2 ) +  2 v_NS_{N-1,2}^2}
\no\\
&&e^{\beta h\sum_{i=1}^{2^N} h_i \sigma_i   -v_N2^{2N}\frac{1}{c \beta^d b_N}} ] 
\no\\
&=&\mathbb{E}[\log\sum_{k=1}^{r_N} \Tr_{ \{ S_{N-1,1}\in I_k\}}   e^{-\beta H_{N-1}^{}(\vec{\sigma}_1 )+ 2 v_NS_{N-1,1}^2 +\beta h\sum_{i=1}^{2^{N-1}} h_i \sigma_i}
\no\\
&&\Tr_{ \{ S_{N-1,2}\in I_k\}}  e^{-\beta H_{N-1}^{}(\vec{\sigma}_2 ) +  2 v_NS_{N-1,2}^2+\beta h\sum_{i=2^{N-1}+1}^{2^N} h_i \sigma_i    } ] -\frac{1}{c \beta^{d-1} }
\no\\
&=&\mathbb{E}[\log\sum_{k=1}^{r_N}Z_{k,1}(\vec{h}_1)  Z_{k,2}(\vec{h}_2) ] -\frac{1}{c \beta^{d-1} },
\ee
where we used Eq. (\ref{res-S1-S2}) in the last inequality.
Furthermore, the Cauchy-Schwartz inequality yields
\be
&&P_N(v_N,v_{N-1},\cdots,v_1)
\no\\
&\ge&\mathbb{E}[\log (\sum_{k=1}^{r_N}Z_{k,1}(\vec{h}_1) )^2  ] - \mathbb{E}[\log \sum_{k=1}^{r_N}\frac{Z_{k,1}(\vec{h}_1)  } { Z_{k,2}(\vec{h}_2) }   ] -\frac{1}{c \beta^{d-1} }
\no\\
&=&2\mathbb{E}[\log  \Tr e^{-\beta H_{N-1}^{\mathrm{RF}}(\vec{\sigma}_1)    + 2v_N S_{N-1,1}^2 }   ] - \mathbb{E}[\log \sum_{k=1}^{r_N} \frac{Z_{k,1}(\vec{h}_1)  } { Z_{k,2}(\vec{h}_2) }  ] -\frac{1}{c \beta^{d-1} }
\no\\
&=&2P_{N-1}(2v_N+v_{N-1},v_{N-2},\cdots,v_1) - \mathbb{E}[\log \sum_{k=1}^{r_N} \frac{Z_{k,1}(\vec{h}_1)  } { Z_{k,2}(\vec{h}_2) }   ] -\frac{1}{c \beta^{d-1} }
\no\\
&\ge&2P_{N-1}(2v_N+v_{N-1},v_{N-2},\cdots,v_1) -\log r_N- \mathbb{E}[ \max_k\log   \frac{Z_{k,1}(\vec{h}_1)  } { Z_{k,2}(\vec{h}_2) }   ] -\frac{1}{c \beta^{d-1} }
\no\\
&\ge&2P_{N-1}(2v_N+v_{N-1},v_{N-2},\cdots,v_1) -\log (1+(c \beta^d b_N)^{1/2})- \mathbb{E}[ \max_k\log  \frac{Z_{k,1}(\vec{h}_1)  } { Z_{k,2}(\vec{h}_2) }   ]
\no\\
&& -\frac{1}{c \beta^{d-1} }, \label{P-P}
\ee
where we used Eq. (\ref{lange-r_N}) in the last inequality.
Finally, by combining Eq. (\ref{f-P}) and Eq. (\ref{P-P}), we prove Lemma \ref{lemma3}.
\end{proof}
To obtain a lower bound on $f_N$ using the recurrence relation, we must evaluate the last term on the right-hand side of Lemma \ref{lemma3}.
Concentration inequalities in probability theory~\cite{BLM} overcome this difficulty.
\begin{lemma}[]\label{lemma5}
\be
\mathbb{E}[ \max_k \log  \frac{Z_{k,1}(\vec{h}_1)   } {Z_{k,2}(\vec{h}_2)  }]  
&\le&\beta h 2^{N/2} \log ((1+(c \beta^d b_N)^{1/2}) (1+\sqrt{2\pi e} ) ).
\ee
\end{lemma}
\begin{proof}[ \textbf{Proof of Lemma \ref{lemma5}}]
First, we consider the case where the random field follows a Gaussian distribution $\mathcal{N}(0,1)$.
Note that $g_k=\log ( Z_{k,1}(\vec{h}_1) /Z_{k,2}(\vec{h}_2)  )$, which is a function of all the random fields, is Lipschitz with a constant $L=\beta h 2^{N/2}$, because we find for any $h_i \, (i=1,\cdots, 2^N)$
\be
\left|\frac{\partial g_k}{\partial h_i} \right| \le \beta h.
\ee
Then, the Gaussian concentration of the Lipschitz functions (the Tsirelson-Ibragimov-Sudakov inequality~\cite{BLM}) indicates  that, for any $t>0$
\be
\Pr(g_k- \mathbb{E}[g_k]\ge t) &\le& e^{ - \frac{t^2}{2L^2}}. \label{TIS-ineq}
\ee
For any $\gamma>0$, the following identity holds~\cite{AK}
\be
\mathbb{E}[ e^{\gamma (g_k- \mathbb{E}[g_k])} ]
&=&\gamma \int_{-\infty}^\infty dt e^{\gamma t}\Pr(g_k- \mathbb{E}[g_k]\ge t)  . \label{exp-id}
\ee
Substituting  Eq. (\ref{TIS-ineq}) into Eq. (\ref{exp-id}), we obtain
 \be
\mathbb{E}[ e^{\gamma (g_k- \mathbb{E}[g_k])} ]
&=&\gamma \int_{-\infty}^0 dt e^{\gamma t}\Pr(g_k- \mathbb{E}[g_k]\ge t)  +\gamma \int_{0}^\infty dt e^{\gamma t}\Pr(g_k- \mathbb{E}[g_k]\ge t) 
\no\\
&\le&\gamma \int_{-\infty}^0 dt e^{\gamma t}  +\gamma \int_{0}^\infty dt e^{\gamma t} e^{ - \frac{t^2}{2L^2}}
\no\\
&\le&\gamma \int_{-\infty}^0 dt e^{\gamma t}  +\gamma \int_{-\infty}^\infty dt e^{\gamma t} e^{ - \frac{t^2}{2L^2}}
\no\\
&=& 1+\sqrt{2\pi} \gamma L e^{\frac{\gamma^2L^2}{2}}. \label{exp-error}
\ee
Thus, we have
\be
\mathbb{E}[\max_k  g_k] -\max_k \mathbb{E}[  g_k]
&\le&\mathbb{E}[\max_k (g_k-\mathbb{E}[  g_k])] 
\no\\
&=& \frac{1}{\gamma}\log \exp(\gamma\mathbb{E}[\max_k  (g_k-\mathbb{E}[  g_k])] )
\no\\
&\le& \frac{1}{\gamma} \log \mathbb{E}[e^{  \gamma \max_k(g_k-\mathbb{E}[  g_k])} ] 
\no\\
&\le& \frac{1}{\gamma}\log (\sum_{k=1}^{r_N} \mathbb{E}[e^{  \gamma(g_k-\mathbb{E}[  g_k])} ] )
\no\\
&\le& \frac{1}{\gamma}\log( r_N (1+\sqrt{2\pi} \gamma L e^{\frac{\gamma^2L^2}{2}} ) )
\no\\
&\le& \frac{1}{\gamma}\log( (1+(c \beta^d b_N)^{1/2}) (1+\sqrt{2\pi} \gamma L e^{\frac{\gamma^2L^2}{2}} )) ,
\ee
where we used the Jensen inequality in the second inequality, Eq. (\ref{exp-error}) in the fourth inequality, and Eq.(\ref{lange-r_N}) in the last inequality.
By setting $\gamma=1/L=1/(\beta h 2^{N/2})$, we obtain
\be
&&\mathbb{E}[ \max_k \log  \frac{Z_{k,1}(\vec{h}_1)  } { Z_{k,2}(\vec{h}_2)  }]
\no\\
&\le&   \max_k\mathbb{E}[  \log  \frac{Z_{k,1}(\vec{h}_1)  } { Z_{k,2}(\vec{h}_2)  }]  +\beta h 2^{N/2}\log ((1+(c \beta^d b_N)^{1/2}) (1+\sqrt{2\pi e}  ) )
\no\\
&=&   \beta h 2^{N/2}\log( (1+(c \beta^d b_N)^{1/2}) (1+\sqrt{2\pi e}  )) , 
\ee
which is the proof of Lemma \ref{lemma5} when the random field follows a Gaussian distribution $\mathcal{N}(0,1)$. 

Finally, we consider the case where the random field follows a Bernoulli distribution.
Note that $h_i$ takes $\pm1$ with a probability of 1/2 and $|g_k(h_i=1)-g_k(h_i=-1)|\le 2\beta h \equiv s$.
Subsequently, the bounded difference inequality (the McDiarmid inequality~\cite{BLM}) indicates that
\be
\Pr( g_k- \mathbb{E}[g_k]\ge t) &\le& e^{ - \frac{2t^2}{2^N s^2}}= e^{ - \frac{t^2}{2L^2}},
\ee
which coincides exactly with Eq. (\ref{TIS-ineq}).
Thus, the calculations are the same.
\end{proof}

We can now prove Theorem \ref{theorem1}.
\begin{proof}[\textbf{Proof of Theorem \ref{theorem1}}]
Lemmas \ref{lemma3} and \ref{lemma5} provide
\be
f_N  &\ge& f_{N-1}- \frac{1}{\beta b_N} \qty(\frac{1}{c \beta^{d-1}} + \log (1+(c \beta^d b_N)^{1/2}) + \beta h 2^{N/2}\log ((1+(c \beta^d b_N)^{1/2}) (1+\sqrt{2\pi e}  ))  ).
\no\\
\ee
As $f_0=1$, we have
\be
f_N &\ge& 1-\sum_{p=1}^\infty \frac{1}{\beta b_p} \qty(\frac{1}{c \beta^{d-1}} + \log (1+(c \beta^d b_p)^{1/2}) + \beta h 2^{p/2}\log ((1+(c \beta^d b_p)^{1/2}) (1+\sqrt{2\pi e} )  )  ).
\no\\
\ee
When $h\neq0$, this infinite series converges to a finite value for any finite $\beta$ and $c$, if $1<\alpha<3/2$.
To avoid the divergence of the logarithmic function at zero temperature ($\beta\to\infty$), we choose $d=0$.
Then, for  $1<\alpha<3/2$ and $1\le(c 2^{(2-\alpha)})^{1/2}=(c  b_1)^{1/2}$, we have
\be
f_N &\ge& 1-\sum_{p=1}^\infty \frac{1}{\beta b_p} \qty(\frac{\beta}{c } + \log (2(c  b_p)^{1/2}) + \beta h 2^{p/2}\log (2(c  b_p)^{1/2}) (1+\sqrt{2\pi e}   )  )
\no\\
&=&  1- \frac{1}{c}\frac{2^\alpha}{2^2-2^\alpha} - h  \left( \frac{(2-\alpha) 2^{1/2+\alpha}}{(2^{3/2}-2^{\alpha})^2}\log  2 + \frac{2^{\alpha}}{2^{3/2}-2^{\alpha}}\log( 2c ^{1/2}(1+\sqrt{2\pi e}   )) \right)
\no\\
&&- \frac{1}{\beta} \left(\frac{2^{\alpha}}{2^2-2^{\alpha}} \log (2c^{1/2}) + \frac{(2-\alpha) 2^{1+\alpha}}{(2^{2}-2^{\alpha})^2}\log  2\right),
\ee
which proves Theorem \ref{theorem1}.
\end{proof}

\section{Discussions}\label{sec4}
We proved the existence of a long-range order at low temperatures, including zero temperature, in the random-field Ising model on the Dyson hierarchical lattice when $1<\alpha<3/2$.
Using concentration inequalities, we extend Dyson's method from the absence of a random field~\cite{Dyson} to the presence of a random field.

When $h=0$, by setting $c=1$ and $d=1$, our proof (Lemma \ref{lemma3}) is equivalent to Dyson's method in the absence of a random field~\cite{Dyson}.
We can then reproduce the existence of a long-range order in the Ising model on the Dyson hierarchical lattice when $0<\alpha<1$~\cite{Dyson}.
However, in the presence of a random field ($h\neq0$), we must choose $c>1$ and $d=0$ to prove the existence of a long-range order at zero temperature.

Our proof is valid only for sufficiently small values of $h$. For sufficiently large $h$, the ferromagnetic order is disrupted by randomness, and the system belongs to the paramagnetic phase at any temperature (for example, see Ref. \cite{DPR} for the phase diagram).

Investigating the one-dimensional long-range random field Ising model in the interval $1<\alpha\le3- \log 3/\log 2$ is an important future problem.
A previous study proved the existence of a long-range order for $3- \log 3/\log 2<\alpha<3/2$~\cite{COP}, but not for $1<\alpha\le3- \log 3/\log 2$.
As the random-field Ising model on the Dyson hierarchical lattice mimics the properties of the one-dimensional long-range random-field Ising model, our results strongly support the existence of a phase transition for $1<\alpha\le3- \log 3/\log 2$ in the one-dimensional long-range random-field Ising model (note that this is not a proof).

\backmatter

%
%
%

\bmhead{Acknowledgements}
This study was supported by JSPS KAKENHI Grant Nos. 24K16973 and 23H01432.
Our study received financial support from the Public\verb|\|Private R\&D Investment Strategic Expansion PrograM (PRISM) and programs for Bridging the gap between R\&D and the IDeal society (society 5.0) and Generating Economic and social value (BRIDGE) from the Cabinet Office.

\section*{Declarations}


\subsection*{Data availability}
No datasets were generated or analyzed in the current study.

\subsection*{Conflict of interest}
The authors declare no conflict of interest.

\bibliography{main-ver6-jstat-revise2}

\end{document}